# Kramers escape rate in overdamped systems with the power-law distribution

Zhou Yanjun, Du Jiulin

*Department of Physics, School of Science, Tianjin University, Tianjin 300072, China*

**Abstract**：Kramers escape rate in the overdamped systems with the power-law distribution is studied. By using the mean first passage time, we derive the escape rate for the power-law distribution and obtain the Kramers' infinite barrier escape rate in this case. It is shown that the escape rate for the power-law distribution extends the Kramers' overdamped result to the relatively low barrier. Furthermore, we apply the escape rate for the power-law distribution to the unfolding of titin and show a better agreement with the experimental rate than the traditional escape rate.

## 1. Introduction

Kramers' problem proposed for the thermal escape of a Brownian particle out of a metastable well has received great attentions and interests in physics, chemical reaction rate, and biology [1,2] etc, which has become the most important one of modern reaction rate theories. At first, Kramers realized a very unsatisfactory feature of transition state theory (TST) which appears to predict escape in absence of coupling to a heat bath in contradiction to the fluctuation-dissipation theorem, and then chose the prefactor $\mu$ to remedy this defect [3]. According to the very low and intermediate-to-high dissipative coupling to the bath, he yielded three explicit formulas of $\mu$ for the escape rate in very low damping, intermediate and overdamped cases, respectively. In this paper, we focus on the overdamped systems.

When Brownian particles move in an overdamped media, the probability distribution of particles can be described by Smoluchowski equation. Kramers postulated that the barrier Brownian particles escape is high compared with thermal energy and the dissipative coupling to the bath is strong so that the systems can be in a thermal equilibrium state, a Maxwell-Boltzmann(MB) distribution always holds in the whole time, and any disturbance to the MB distribution can almost be negligible at all times [3]. Under this assumption, the escape rate is obtained. However, the assumption is farfetched in open complex systems. Nonequilibrium is the main feature of an open complex system. In fact, lots of experimental studies on complex systems have shown non-MB distributions or power-law distributions, such as glasses [4, 5], disordered media [6-8], folding of proteins [9], single-molecule conformational dynamics [10, 11], trapped ion reactions [12], chemical kinetics, and biological and ecological population dynamics [13, 14], reaction–diffusion processes [15], chemical reactions [16], combustion processes [17], gene expression [18], cell reproductions [19], complex cellular networks [20], small organic molecules [21],and astrophysical and space plasmas [22], etc. The typical forms of such power-law distributions have included the $\kappa$-distributions or the generalized Lorentzian distributions in the solar wind and space plasmas [22-26], the $q$-distributions in complex systems within nonextensive statistical mechanics [27], and those $\alpha$-distributions noted in physics, chemistry and elsewhere like $P(E) \sim E^{-\alpha}$ with an index $\alpha > 0$ [12, 15, 16, 21, 23, 28].

These power-law distributions may lead to processes different from those in the realm governed by Boltzmann–Gibbs statistics with MB distribution. Simultaneously, a class of statistical mechanical theories studying the power-law distributions in complex systems has been constructed, for instance, by generalizing Boltzmann entropy to Tsallis entropy [27], by generalizing Gibbsian theory [29] to a system away from thermal equilibrium, and so forth. Most recently, a generalized transition state theory (TST) for the systems with power-law distributions was studied and the generalized reaction rate formulae were presented for one-dimensional and $n$-dimensional Hamiltonian systems away from equilibrium [30]. And the power-law TST reaction rate coefficient for an elementary bimolecular reaction was also studied when the reaction takes place in the system with power-law distributions [31]. These developments now naturally give rise to a possibility that Kramers escape rate can be reconsidered under the condition of power-law distributions in a complex system.

One of the most important thermal escape theories is the first passage time theory [32, 33]. In an overdamped case, suppose the motions of the particles are bounded in the finite space $V$ with an absorbing boundary $\Sigma$, and they are governed by the Langevin equation [34],

$$\frac{dx}{dt} = -(m\gamma)^{-1} \frac{dU}{dx} + (m\gamma)^{-1} \eta(x,t), \tag{1}$$

where $m$ is the mass of the particle, $U$ is a potential field, $\gamma$ is the friction coefficient. Generally, the noise may be considered inhomogeneous in space and so it may be a function of the variables $x$, i.e. $\eta = \eta(x,t)$ is a multiplicative (space-dependent) noise. It is usually assumed that the noise is Gaussian, with zero average and delta-correlated in time $t$, such that it satisfies,

$$\langle \eta(x,t) \rangle = 0, \quad \langle \eta(x,t)\eta(x,t') \rangle = 2D(x)\delta(t-t'). \tag{2}$$

The correlation strength of multiplicative noise (i.e. diffusion coefficient) $D(x)$ is also a function of the variables $x$.

If $P(x,t)$ is the probability distribution of particles that has not left by time $t$, it satisfies the Smoluchowski equation,

$$\frac{\partial P(x,t)}{\partial t} = (m\gamma)^{-1} \frac{\partial}{\partial x}\left( \frac{dU}{dx} P(x,t) \right) + (m\gamma)^{-1} \frac{\partial}{\partial x}\left( D(x) \frac{\partial P(x,t)}{\partial x} \right), \tag{3}$$

where the initial condition and absorbing boundary condition are $P(x,0)=\delta(x-x_0)$, $P(x,t)\big|_{\Sigma} = 0$ respectively. It is known that when the system reaches thermal equilibrium, $D(x)$ is a constant, $D=\beta^{-1}$ and the stationary solution of Eq.(3) is an MB distribution; When the system is away from thermal equilibrium, one can consider $D(x)$ as a function of the potential energy $U(x)$, and to satisfy the generalized fluctuation-dissipation relation (FDR) for power-law distribution [34],

$$D(x) = \beta^{-1}\left[1 - \kappa\beta U(x)\right], \quad (4)$$

where $\beta = 1/k_B T$, $k_B$ is the Boltzmann constant, $T$ is the temperature, $\kappa$ is the power-law parameter and $\kappa \neq 0$ measures a distance away from thermal equilibrium. Substituting Eq. (4) into Eq. (3), one can show that the stationary-state solution is the power-law $\kappa$-distribution [34],

$$P_s(x) = Z_\kappa^{-1}\left[1 - \kappa\beta U(x)\right]_+^{1/\kappa}, \quad (5)$$

where $Z_\kappa$ is the normalization constant $Z_\kappa = \int dx \left[1 - \kappa\beta U(x)\right]_+^{\frac{1}{\kappa}}$. It is clear that the generalized FDR is a condition under which the power-law $\kappa$-distribution can be created from the stochastic dynamics of the Langevin equations. At the same time, we can easily find that Eq.(3) combined with Eq.(4) lead to the following time-dependent solution of the power-law distribution,

$$P(x,t) = Z_\kappa^{-1}(t)\left[1 - \kappa\beta(t)(x - x_M(t))^2\right]_+^{\frac{1}{\kappa}}, \quad (6)$$

where $Z_\kappa(t) = \int dx \left[1 - \kappa\beta(t)(x - x_M(t))^2\right]_+^{\frac{1}{\kappa}}$ is the normalization constant at time $t$, $\beta(t) = 1/k_B T(t)$, $T(t)$ is the temperature at time $t$, $x_M(t)$ is the sym-center of the potential energy at time $t$. The time-dependent solution is precisely the same as the result with $q$-distributions (see Eq. (9) in Ref. [35]).

We introduce the Fokker-Planck operator, $f$, and its adjoint operator, $f^\dagger$ [32],

$$f = (m\gamma)^{-1}\frac{\partial}{\partial x}\left(\frac{dU}{dx}\right) + (m\gamma)^{-1}\frac{\partial}{\partial x}D(x)\frac{\partial}{\partial x}, \quad (7a)$$

$$f^\dagger = -(m\gamma)^{-1}\frac{dU}{dx}\frac{\partial}{\partial x} + (m\gamma)^{-1}\frac{\partial}{\partial x}D(x)\frac{\partial}{\partial x}, \quad (7b)$$

and as an initial value problem, the operator solution is $P(x,t) = e^{tf}\delta(x - x_0)$. Integrating $P(x,t)$ over the volume $V$, the numbers of all starting points $s(t, x_0)$ that are still in $V$ at time $t$ are obtained, which depends on the initial value $x_0$. The number difference of the initial points that have not left before time $t$ but have left during the time interval $dt$ after time $t$ determines the probability density of lifetime $\rho(t, x_0)$ [32],

$$\rho(t, x_0) = -\frac{d}{dt}s(t, x_0). \quad (8)$$

The first passage time is the first time that the particles leave the boundary $\Sigma$. Because of the noise, even the same initial positions will lead to different first passage time, hence the mean first passage time (MFPT), $\tau(x)$, which is the first moment of $t$, is introduced. There is a more direct way to calculate MFPT. Namely, using the adjoint operator $f^\dagger$ to operate on the MFPT, then the MFPT is determined by solving the inhomogeneous adjoint equation [32],

$$f^{\dagger}\tau(x) = -1, \text{ and } \tau(x) = 0 \text{ on } \Sigma. \qquad (9)$$

If initial positions of the particles are located in the reactant domain and the finite space $V$ is the domain for the reaction, then the transition from the reactant to the boundary may be characterized by a rate which is simply determined by the inverse MFPT [32, 33], $k = 1/2\tau$, where $k$ is the escape rate, the prefactor 1/2 means the equal probability of the forward and backward crossing, so $k$ is half of it. For the low damping systems, the particle that reaches the maximum has enough energy to overcome the barrier and rapidly slides down to the next potential minimum, so the prefactor is 1 [36].

This paper is organized as follows. In Section 2, the transformed operator $f$ and adjoint operator $f^{\dagger}$ are derived with the power-law distribution, and the generalized Kramers' escape rate in an overdamped system is derived. In Section 3, we discuss the finite barrier effects by comparing with the Kramers' escape rate in Section 2 (i.e. Kramers' infinite barrier result). In the case of Josephson junctions for instance [36], where the overdamped regime can be addressed with resistively shunted tunnel junctions or with superconducting-normal-superconducting, the barrier height is $\frac{\gamma_0}{2}\left[-\pi\alpha + 2\left(\alpha\sin^{-1}\alpha + \sqrt{1-\alpha^2}\right)\right]$, where $\gamma_0 = \Phi_0 I_C / \pi k_B T$, $\alpha \equiv I_0 / I_C$, $\Phi_0$ is the magnetic flux quantum, $I_0$ is the bias current of the circuit and $I_C$ is the critical current. We see that the barrier decreases with the increasing of $\alpha$, and there is zero barrier height when $\alpha = 1$. Take the pulling experiment of single-molecular as another example [37, 38]; the barrier height is $\Delta G$ in the absence of the external force $F$. When $F$ is applied to the molecular, the barrier height decreases with the increasing $F$. Therefore, research on the finite barrier effect is important. In Section 4, we take the experiment of unfolding of titin as an example and apply our theoretical result to it. Finally in Section 5, the conclusion is given.

## 2. The overdamped Kramers' escape rate with the power-law distribution

2.1 The general expression of escape rate with the power-law distribution

The Smoluchowski equation, Eq. (3), with the FDR Eq.(4) can be written equivalently as

$$\frac{\partial P(x,t)}{\partial t} = (m\gamma)^{-1} \frac{\partial}{\partial x}\left(D(x)e^{-\phi_\kappa}\frac{\partial e^{\phi_\kappa}P(x,t)}{\partial x}\right), \qquad (10)$$

where $\phi_\kappa$ is defined as a potential with the power-law distribution, $\phi_\kappa = -\ln(1-\kappa\beta U)^{1/\kappa}$, when $\kappa \to 0$, $\phi_\kappa$ returns to the usual form $\phi = \beta U$. Consequently, the transformed operator $f_\kappa$ and adjoint operator $f_\kappa^{\dagger}$ become respectively,

$$f_\kappa = (m\gamma)^{-1} \frac{\partial}{\partial x} e^{-\phi_\kappa} D(x) \frac{\partial}{\partial x} e^{\phi_\kappa}, \qquad (11a)$$

$$f_\kappa^\dagger = (m\gamma)^{-1} e^{\phi_\kappa} \frac{\partial}{\partial x} e^{-\phi_\kappa} D(x) \frac{\partial}{\partial x}. \qquad (11b)$$

Take Eq. (11$b$) into the first equation in Eq. (9) and the partial differential equation of $\tau$ with absorbing boundary is established by

$$(m\gamma)^{-1} e^{\phi_\kappa} \frac{\partial}{\partial x} e^{-\phi_\kappa} D(x) \frac{\partial \tau(x)}{\partial x} = -1. \qquad (12)$$

Supposing the coordinate $x$ is the starting position of the Brownian particle, the absorbing barrier is located at $b$, and there is a reflecting barrier at $a$, with $a < x < b$. To solve the equation, we divide Eq. (12) by $e^{\phi_\kappa}$, integrate once over $x$, multiply by $e^{\phi_\kappa}/D(x)$, integrate once more over $x$, and use the boundary conditions at the barrier $\tau(x=b)=0$, and then we get the general result of escape rate with the power-law distribution,

$$k_\kappa = \frac{1}{2\tau(x)} = \frac{1}{2m\gamma \int_x^b dy \frac{e^{\phi_\kappa(y)}}{D(y)} \int_a^y dz\, e^{-\phi_\kappa(z)}}, \qquad (13)$$

Eq. (13) is the key results of this work. Next, we derive the Kramers' escape rate according to Eq. (13).

2.2 The Kramers' escape rate with the power-law distribution

Suppose that the absorbing barrier is placed at the maximum $x_{\max}$ of the potential $U(x)$, and $U_{\max} = U(x_{\max})$. The initial position of particles is $x$, and the reflecting barrier at $x = a$ is provided by a repelling potential at $x \to -\infty$ [32]. Then Eq.(13) becomes

$$k_\kappa = \frac{1}{2m\gamma \int_x^{x_{\max}} dy \frac{e^{\phi_\kappa(y)}}{D(y)} \int_{-\infty}^y dz\, e^{-\phi_\kappa(z)}}. \qquad (14)$$

Assuming the region where the potential energy has a minimum $U_{\min}$, the potential energy is diagonal and expanded as a harmonic function [32],

$$U(z) = U_{\min} + \frac{1}{2}\omega_{\min}^2 (z - x_{\min})^2 + \ldots, \qquad (15a)$$

the potential energy has a maximum $U_{\max}$ which is located at $x_{\max}$, and it is written as

$$U(y) = U_{\max} - \frac{1}{2}\omega_{\max}^2 (y - x_{\max})^2 + \ldots. \qquad (15b)$$

As mentioned in the introduction, if the diffusion coefficient $D$ satisfies Eq.(4), the stationary solution of Eq.(3) is the power-law $\kappa$-distribution. Take $D$ and $\phi$ into the first integration of the denominator in Eq.(14), and postulate the integral over $y$ is practically independent of $x$ as long as $x$ is near the potential minimum, so the lower limit can be replaced by minus infinity [32], hence we obtain

$$\int_x^{x_{max}} dy \frac{e^{\phi_\kappa(y)}}{D(y)} \approx \int_{-\infty}^{x_{max}} dy \frac{e^{\phi_\kappa(y)}}{D(y)} = (1-\kappa\beta U_{max})^{\frac{\kappa+1}{-\kappa}} \int_{-\infty}^{x_{max}} dy \beta \left[1 + \frac{\kappa\beta\omega_{max}^2(y-x_{max})^2}{2(1-\kappa\beta U_{max})}\right]^{\frac{\kappa+1}{-\kappa}},$$
(16)

when thermal energy $k_B T$ is much smaller than the barrier height $U_{max}$-$U_{min}$ (that is also called infinite barrier case). The integral over $z$ is dominated by the potential near the minimum, then the upper limit of integration can be replaced by infinity [32],

$$\int_{-\infty}^y dz e^{-\phi_\kappa(z)} \approx \int_{-\infty}^{\infty} dz e^{-\phi_\kappa(z)} = (1-\kappa\beta U_{min})^{\frac{1}{\kappa}} \int_{-\infty}^{\infty} dz \left[1 - \frac{\kappa\beta\omega_{min}^2(z-x_{min})^2}{2(1-\kappa\beta U_{min})}\right]^{\frac{1}{\kappa}}. \quad (17)$$

For Eq.(16), when $\kappa < 0$, the cutoff condition requires $1 + \frac{\kappa\beta\omega_{max}^2(y-x_{max})^2}{2(1-\kappa\beta U_{max})} > 0$, and so the lower limit of the integral becomes

$$\int_{-\infty}^{x_{max}} dy \frac{e^{\phi_\kappa(y)}}{D(y)} = (1-\kappa\beta U_{max})^{\frac{\kappa+1}{-\kappa}} \int_{x_{max}-\sqrt{\frac{2(1-\beta\kappa U_{max})}{-\beta\omega_{max}^2\kappa}}}^{x_{max}} dy \beta \left[1 + \frac{\kappa\beta\omega_{max}^2(y-x_{max})^2}{2(1-\kappa\beta U_{max})}\right]^{\frac{\kappa+1}{-\kappa}}. \quad (18)$$

For Eq. (17), when $\kappa > 0$, the cutoff condition requires $1 - \frac{\kappa\beta\omega_{min}^2(z-x_{min})^2}{2(1-\kappa\beta U_{min})} > 0$, and so we have

$$\int_{-\infty}^{\infty} dz e^{-\phi_\kappa(z)} = (1-\kappa\beta U_{min})^{\frac{1}{\kappa}} \int_{x_{min}-a}^{x_{min}+a} dz \left[1 - \frac{\kappa\beta\omega_{min}^2(z-x_{min})^2}{2(1-\kappa\beta U_{min})}\right]^{\frac{1}{\kappa}}, \quad (19)$$

with $a = \sqrt{\frac{2(1-\beta\kappa U_{min})}{\beta\omega_{min}^2\kappa}}$. Integrate Eq.(16) and Eq.(17) and take the integral results into Eq.(14), we obtain

$$k_\kappa = \left(1 + \frac{\kappa}{2}\right) \frac{\omega_{min}\omega_{max}}{2\pi m\gamma} \left(\frac{1-\beta\kappa U_{max}}{1-\beta\kappa U_{min}}\right)^{\frac{1}{2}+\frac{1}{\kappa}}, \quad (20)$$

It is clear that in the limit $\kappa \to 0$ the expression of standard Kramers' overdamped escape rate [3, 32] is recovered,

$$k = \frac{\omega_{min}\omega_{max}}{2\pi m\gamma} \exp\left[-\beta(U_{max}-U_{min})\right]. \quad (21)$$

Thus, we obtain a generalized Kramers' overdamped escape rate in one-dimensional nonequilibrium system with the power-law distribution.

## 3. Discussion of the finite barrier effect

In the former calculations, Kramers made an assumption that the barrier height was very larger than the thermal energy. In practice, it is not always right to do so. We have listed several examples in the introduction to illustrate the existence of low barrier height. So the finite barrier effect needs to be considered. Now, we study how the finite barrier effect to affect the escape rate. The relevant parameters are taken as $U_{min}=0$, $U_{max}=\Delta U$, $\omega_{min}=\omega_{max}=1$, $x_{min}=0$, $x_{max}=\sqrt{2\Delta U}$ in Eq.(15$a$) and Eq.(15$b$), where $\Delta U$ is the barrier height. Then Eq.(14) becomes

$$k_\kappa = \frac{1}{2m\gamma\beta \int_x^{\sqrt{2\Delta U}} dy \left[1-\kappa\beta\Delta U + \frac{1}{2}\kappa\beta\left(y-\sqrt{2\Delta U}\right)^2\right]^{\frac{\kappa+1}{-\kappa}} \int_{-\infty}^{y} dz \left(1-\kappa\beta\frac{1}{2}z^2\right)^{1/\kappa}}. \qquad (22)$$

We make variable transforms, $\sqrt{\beta}\,y=Y$ and $\sqrt{\beta}\,z=Z$, thus it becomes

$$k_\kappa = \frac{1}{2m\gamma \int_X^{\sqrt{2\beta\Delta U}} dY \left[1-\kappa\beta\Delta U + \frac{1}{2}\kappa\left(Y-\sqrt{2\beta\Delta U}\right)^2\right]^{\frac{\kappa+1}{-\kappa}} \int_{-\infty}^{Y} dZ \left(1-\frac{1}{2}\kappa Z^2\right)^{1/\kappa}}. \qquad (23)$$

The numerical results of Eq.(23), the escape rate in the overdamped limit normalized by the power-law Kramers result in Eq.(20) for different $\beta\Delta U$, are shown in Fig.1.

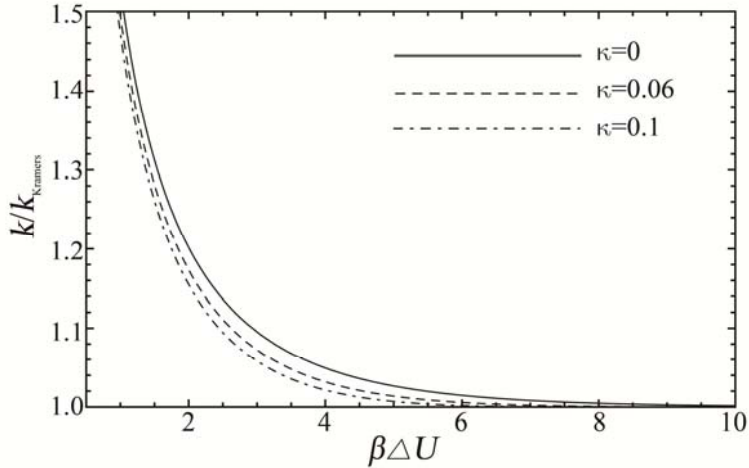

Fig.1 The escape rate in the overdamped case with the power-law distribution.

Fig.1 gives a quantitative estimation of the accuracy of the Kramers overdamped result for every barrier height compared with the thermal energy. It is shown that if $\beta\Delta U$ is very small, the curves for the ratio $k_\kappa/k_{Kramers}$ with three different values of the $\kappa$-parameter almost overlapped together. Because, if $\beta\Delta U$ is very small, there is almost no difference between the function $e^x$ and the function $e^x_\kappa = (1-\kappa x)^{1/\kappa}$, and the ratio $k_\kappa/k_{Kramers}$ depends only on $\beta\Delta U$. With $\beta\Delta U$ increasing, $k_\kappa/k_{Kramers}$ becomes small. The bigger the power-law parameter is, the greater $k_\kappa/k_{Kramers}$ approaching to 1 is. In the normal case, i.e. $\kappa = 0$, $k_\kappa/k_{Kramers} = 1.095$ at $\beta\Delta U=3$ and $k_\kappa/k_{Kramers} = 1.026$ at $\beta\Delta U=5$; whereas in the case $\kappa \neq 0$, take $\kappa = 0.1$ for instance, $k_\kappa/k_{Kramers} = 1.058$ at $\beta\Delta U = 3$ and $k_\kappa/k_{Kramers} = 1.007$ at $\beta\Delta U = 5$, all of which are much closer to 1 than that for $\kappa=0$ case. Thus, the Kramers' overdamped result with the power-law distribution has a wider application in the range of relatively low $\beta\Delta U$ if the high accuracy is considered.

**4. Combination of theoretical results with the experiments**

There exist plenty of real systems describing the very high damping condition, such as a biased Josephson junction [36], single-molecule pulling experiments in biology [37, 38] and the atomic friction [39, 40] etc. Here we apply the generalized expression of escape rate, Eq.(20), to the pulling experiment in biology.

In section 2, we derived the Kramers escape rate in the harmonic potential. Here for the specific system under consideration, the potential has another forms, such as the pulling experiment of unfolding of titin where the cusp potential is used (see Eq.(24) below), the periodic potential $U(x) = U_0(1-\cos x) - Fx$ in the biased Josephson junction [36] where $U_0$ is the bare potential energy and $F$ is the external force, the periodic potential $U(R,x) = -U_0 \cos(2\pi x/a) + k_0[R(t)-x]^2/2$ in the atomic friction [39] where $a$ is the lattice constant, $k_0$ is the elastic constant, $R(t)$ is the dragging coordinate and $U_0$ is the surface barrier potential height.

For the pulling experiment, the molecule moves on a combined free-energy surface. This free-energy [37] includes the bare free-energy $U_0(x)$ and the approximately elastic potential energy of soft spring $-F(t)x$ along the pulling coordinate $x$, $U(x) = U_0(x) - F(t)x$. The bare free-energy $U_0(x)$ is

$$U_0(x) = \begin{cases} \Delta G \left(\dfrac{x}{x_{\neq}}\right)^2, & x < x_{\neq} \\ -\infty, & x \geq x_{\neq} \end{cases} \quad (24)$$

where $x_{\neq}$ denotes the transition state, $\Delta G$ is the barrier height at $x = x_{\neq}$, the external force $F$ is ramped up linearly with time, $F = K\upsilon t$, with the spring constant $K$ and the pulling velocity $\upsilon$. One can predict the escape rate $k$ in the presence of a force $F$ using Kramers' theory as follows,

$$\frac{k}{k_0} = \frac{\int\limits_{barrier} dy \frac{e^{\phi(y)}}{D(y)} \int\limits_{well} dz e^{-\phi(z)}}{\int\limits_{barrier} dy \frac{e^{\phi'(y)}}{D(y)} \int\limits_{well} dz e^{-\phi'(z)}}, \quad (25)$$

where $k_0$ is the bare escape rate, $\phi'$ and $\phi$ are the relevant potentials with the power-law distribution, given by $\phi' = -\ln[1-\kappa\beta(U_0-Fx)]^{1/\kappa}$ and $\phi = -\ln[1-\kappa\beta U_0]^{1/\kappa}$, the integrals extend over the well and barrier regions respectively. Thus, $k_0$ is

$$k_0 = \frac{\Delta G(1-\kappa\beta\Delta G)^{\frac{1}{\kappa}}}{m\gamma x_{\neq}^2} \begin{cases} \sqrt{\frac{\kappa\beta\Delta G}{\pi}} \Gamma\left(\frac{1}{\kappa}+\frac{3}{2}\right) / \Gamma\left(\frac{1}{\kappa}+1\right), & \kappa > 0 \\ \sqrt{\frac{-\kappa\beta\Delta G}{\pi}} \Gamma\left(-\frac{1}{\kappa}\right) / \Gamma\left(-\frac{1}{\kappa}-\frac{1}{2}\right), & -2 < \kappa < 0 \end{cases} \quad (26)$$

In the pulling experiment, there is usually an external force $F$ involved in it. But more attentions are paid how to extract reliable information about the kinetics of molecular transitions in the absence of external forces [37], such as the intrinsic rate $k_0$. The experimental process is carried as follows: the pulling of molecular can be considered as an escape process of crossing the barrier. Since the escape process is stochastic and the external force of each escape is also stochastic, which result in the distribution function $P(F)$ measured in the experiment. According to the relationship between $P(F)$ and escape rate $k$ [36, 37, 41], $P(F) = \frac{k(F)}{\dot{F}} \exp\left[-\int_0^F \frac{k(F')}{\dot{F}} dF'\right]$ as well as $k$ and $k_0$, other parameters are obtained by fitting the observed distribution function $f(F)$, such as the intrinsic rate $k_0$, and the transition state $x_{\neq}$ etc.

In the pulling experiment of unfolding of titin, the material is a recombinant construct of 8 tethered I27 titin molecules, and the unfolding of multiple covalently linked proteins can be treated as the unfolding of a single protein with an average effective pulling spring constant. The apparent rate $\tilde{k}_0$ is roughly between one and eight times than the intrinsic rate $k_0$ of a single titin, and the specific value of $k_0$ is confirmed according to the rupture event of molecules in the experiment [38]. Here we choose one situation to further discuss, i.e. the 8-th rupture event happens, thus $\tilde{k}_0 = k_0$; for other cases, such as the $i$-th rupture event happens ($1 \leq i \leq 8$), then $k_0 = \tilde{k}_0/(8-i+1)$. The distribution function was observed for the pulling velocity $\upsilon = 0.6\mu m \cdot s^{-1}$ and the experimental intrinsic rate was $k_0 = 5 \times 10^{-4}$, the barrier height was $\Delta G = 19.3 k_B T$, barrier position was $x_{\neq} = 0.42 nm$, $T = 298K$ [42], $m\gamma = 4.6 \times 10^{-5} kg \cdot s^{-1}$ (this value is got according to $D = k_B T / m\gamma$), the theoretical expression of $k_0$ (see Eq. (11) in Ref.[38]), $\Delta G = 19.3 k_B T$, $x_{\neq} = 0.42 nm$, $T = 298 K$ and $k_0 = 10^{-4}$). We take above quantities into Eq. (26) and find that the theoretical result of $k_0$ is good agreement with the experimental one if the power-law parameter take $\kappa = -0.0097$. If the other rupture event occurs instead of the case we have chosen above, such as $i=7$, the Hummer's predicted value $k_0 = 0.5 \times 10^{-4}$ is much smaller than the experimental one, but our result still keeps the same as the experimental data when the power-law parameter take $\kappa = -0.0147$. Thereby, we have shown an expected result by using the generalized escape rate. Dudko *et al* proposed a unified formalism of model-independent for extracting $k_0$, $x_{\neq}$ and $\Delta G$ from the pulling experiments, and introduced an additional fitting parameter $\nu$ by the mathematical method [37]. Whereas we established an expression of the instinct rate under the power-law $\kappa$-distribution. Hence, our theory with the power-law distribution excellently represented the real systems away from equilibrium.

## 5. Conclusion

In the traditional Kramers escape rate theory, where Brownian particles move in an overdamped homogeneous media, the diffusion coefficient is a constant and the stationary solution of Smoluchowski equation is a MB distribution. For sufficiently high barriers, the dissipative coupling to the bath is so strong that a MB distribution always holds in the whole escape process. However, if the media in the systems is inhomogeneous, the diffusion coefficient may be a function of the space variable $x$. For example, if the generalized FDR in overdamped complex systems is given [34] by $D(x) = \beta^{-1}[1 - \kappa \beta V(x)]$, the stationary solution of Smoluchowski equation is the power-law $\kappa$-distribution and therefore for such systems the Kramers escape rate need to be generalized. In this work, according to the first passage time theory the Kramers escape rate has been restudied for the systems with the power-law $\kappa$-distribution. Because of the noise, even the same initial positions will lead to different first passage times, and hence the MFPT is introduced, which connects with the escape rate by solving the partial differential equation that MFPT satisfies. Thus, the generalized expressions of MFPT and the relevant escape rate are derived for the $\kappa$-distribution.

We then discussed the finite barrier effect and compared it with the Kramers' infinite barrier result. For the potential barrier, the Kramers' assumption that the barrier height is much larger than thermal energy is not valid all the time for the systems in biology, and in super-conduction etc. Therefore, it is necessary to study the finite barrier effect. We showed that the escape rate with the power-law $\kappa$-distribution extends the Kramers' overdamped result to relatively low barrier height. Furthermore, we applied our theoretical result to the pulling experiment of unfolding of titin. In the case of the cups potential and with the power-law $\kappa$-distribution, we calculated the instinct rate $k_0$. In the experiment, the escape process is stochastic and the external force *F* of the escape is also stochastic, thus the distribution function of *F* is observed. According to the relationship between the distribution function of *F* and the escape rate *k*, as well as *k* and $k_0$, the instinct rate $k_0$ can be obtained. It is shown that the theoretical result of the rate for the power-law distribution is good agreement with the experimental result. It is expected that the generalized Kramers escape rate can lead to an insight into the research on the reaction rate theory for the nonequilibrium complex systems with power-law distributions.

**Acknowledgments**
This work is supported by the National Natural Science Foundation of China under grant No 11175128 and by the Higher School Specialized Research Fund for Doctoral Program under grant No 20110032110058.